\documentclass[10pt,twocolumn,letterpaper]{article}

\usepackage{cvpr}      
\usepackage{booktabs}
\usepackage{adjustbox}   
\usepackage{caption} 
\usepackage{siunitx}      
\newcommand{\red}[1]{{\color{red}#1}}

\newcommand{\orange}[1]{\textcolor{orange}{#1}}

\newcommand{\methodName}{Conformable Convolution}

\usepackage{cuted} 
\usepackage{multirow}
\newcommand{\xpm}[1]{{\tiny$\pm#1$}}
\usepackage{amsthm}

\newtheorem*{hyp*}{Hypothesis \protect\hypnumber} 

\newcommand{\hypnumber}{}

\definecolor{cvprblue}{rgb}{0.21,0.49,0.74}
\usepackage[pagebackref,breaklinks,colorlinks,allcolors=cvprblue]{hyperref}

\def\paperID{10553} 
\def\confName{CVPR}
\def\confYear{2025}

\title{Conformable Convolution for Topologically Aware Learning of Complex Anatomical Structures}

\author{Yousef Yeganeh$^{1,2}$ \and
Rui Xiao$^{1}$ \and
Goktug Guvercin$^{1}$ \and
Nassir Navab$^{1,2}$ \and
Azade Farshad$^{1,2}$ \and \\ 
$^{1}$Technical University of Munich \\
$^{2}$Munich Center of Machine Learning \\}

\begin{document}
\maketitle
\begin{abstract}
While conventional computer vision emphasizes pixel-level and feature-based objectives, medical image analysis of intricate biological structures necessitates explicit representation of their complex topological properties. Despite their successes, deep learning models often struggle to accurately capture the connectivity and continuity of fine, sometimes pixel-thin, yet critical structures due to their reliance on implicit learning from data. Such shortcomings can significantly impact the reliability of analysis results and hinder clinical decision-making. To address this challenge, we introduce \textbf{\methodName{}}, a novel convolutional layer designed to explicitly enforce topological consistency. \methodName{} learns adaptive kernel offsets that preferentially focus on regions of high topological significance within an image. This prioritization is guided by our proposed \textbf{Topological Posterior Generator} (TPG) module, which leverages persistent homology. The TPG module identifies key topological features and guides the convolutional layers by applying persistent homology to feature maps transformed into cubical complexes. Our proposed modules are architecture-agnostic, enabling them to be integrated seamlessly into various architectures. We showcase the effectiveness of our framework in the segmentation task, where preserving the interconnectedness of structures is critical. Experimental results on three diverse datasets demonstrate that our framework effectively preserves the topology in the segmentation downstream task, both quantitatively and qualitatively.
\end{abstract}
\section{Introduction}
Recent advances in medical image analysis, particularly in segmentation\cite{farshad2022metamedseg,yeganeh2023transformers,roy2023few,farshad2023learning,yeganeh2023anatomy,yeganeh2020inverse,mozafari2023visa,zerouaoui2024amonuseg}, have often prioritized pixel-level accuracy or visual quality, neglecting the inherent topological properties of anatomical structures. This oversight can lead to critical topological errors like false splits, merges, holes, or disconnected components, compromising the accuracy and reliability of analyses with potentially severe clinical consequences. For example, failing to accurately detect a ruptured vessel may lead to misdiagnosis of conditions like aneurysms or stenoses. Therefore, ensuring realistic topological coherence is paramount in medical image analysis, where the continuity and connectivity of structures like vessels are essential. While SOTA models \cite{hatamizadeh2021swin,he2023swinunetr,zhou2021study} demonstrate strong performance on pixel-wise metrics, they often fail to capture these crucial topological characteristics.

To address this gap, we introduce \methodName{}, an adaptive convolutional layer that explicitly incorporates topological priors into the learning process, enhancing the model's ability to capture topologically relevant features. The \methodName{} layers dynamically adjust sampling locations within their receptive field through learnable offsets, enabling the model to focus on regions of high topological interest. To identify these regions, we propose a novel Topological Posterior Generator (TPG) module that leverages persistent homology \cite{edelsbrunner2002topological} to quantify topological features across different scales – from connected components to loops and voids. By applying persistent homology to cubical complexes derived from feature maps, we obtain a discrete representation that effectively captures the underlying topology. \methodName{} layers are architecture-agnostic and seamlessly replace standard convolutions within existing architectures. This makes them easy to integrate into various models to enforce topological preservation across diverse medical image analysis tasks, including segmentation.

We evaluate our framework on three diverse medical imaging datasets, where the continuity and connectivity of the structures are essential. Our framework effectively adheres to the topology in the input images, improving segmentation performance both qualitatively and quantitatively through conventional pixel-level segmentation metrics as well as connectivity-based metrics. The results of our evaluation on CHASE\_DB1 \cite{fraz2012ensemble} for retinal vessel segmentation, HT29 \cite{carpenter2006cellprofiler,ljosa2012annotated} for colon cancer cell segmentation, and ISBI12 \cite{arganda2015crowdsourcing} for neuron electron microscopy (EM) segmentation, demonstrate the effectiveness of the proposed modules in different shapes and structures. Furthermore, we propose a new evaluation metric through blood flow simulation to show the effectiveness of our model in vascular structures, which is presented in the supplementary materials. 

To summarize our main contributions: (1) We propose \methodName{}, which are convolutional layers with adaptively adjustable kernels guided by topological priors; (2) We propose the Topological Posterior Generator (TPG) module, which extracts the topological regions of interest for guiding the \methodName{}, (3) Our proposed modules are architecture-agnostic and can replace any convolution-based layer, (4) The quantitative and qualitative results of our experiments on the segmentation downstream task on three different organs and structures demonstrate the high impact of the proposed modules in topological metrics while achieving comparable or higher performance in pixel level metrics.
\section{Related Works}
Previous work on topology-preserving methods can be broadly categorized into topology-aware networks and topology-aware objective functions \cite{wasserman2018topological}. 
In addition, we cover methods that are not necessarily developed to preserve topological structures but are relevant to our design.

\paragraph{Topology-preserving Layers and Networks} Hofer \etal
\cite{hofer2017deep} designed an input layer for a network that enables topological signatures as the input and learning the optimal representations during training. \cite{wyburd2021teds} utilizes the transformer-based VoxelMorph \cite{balakrishnan2019voxelmorph} framework that learns to deform a topologically correct prior into the actual segmentation mask. However, such a method could merely deform complex shapes such as vessels. Besides, Yeganeh \etal \cite{yeganeh2023scope} proposes a graph-based method that preserves continuity in retinal image segmentation. Wang \etal \cite{wang2022ta} introduces a topology-aware network and utilizes medial axis transformation to encode the morphology of densely clustered gland cells in histopathological image segmentation. Gupta \etal \cite{gupta2022learning} employed a constraint-based approach to learn anatomical interactions, thereby facilitating the differentiation of tissues in medical segmentation. Horn \etal \cite{horn2021topological} presents a topological layer into Graph Neural Networks. Gupta \cite{gupta2024topology} employs Discrete Morse Theory (DMT) \cite{forman2002user} for structural uncertainty estimation in Graph Convolution Networks (GCN) \cite{kipf2016semi}. Nishikawa \cite{nishikawa2024adaptive} applies Persistent Homology for point cloud analysis. Yi \cite{yi2024learning} proposes geometric-ware modeling for topology preservation in scalp electroencephalography (EEG). Moor \etal \cite{moor2020topological} constrains the bottleneck layer of an Autoencoder to produce topologically correct features. Similar to their method, our method could most effectively be applied to the bottleneck to produce highly topological faithful features.

\paragraph{Topology-preserving Objectives} ToPoLoss \cite{hu2019topology,clough2020topological} minimizes the Wasserstein distance in the \textit{persistence diagram} \cite{vaserstein1969markov,cohen2010lipschitz} between the prediction and the ground truth. Stucki \cite{stucki2023topologically} further improves such a Wasserstein matching by adopting the induced matching method on the persistence barcodes. Prior to that, Centerline Dice (clDice) \cite{shit2021cldice} was proposed as a tubular-structure-dedicated metric and loss function that improves the segmentation results with accurate connectivity information. Another topology-aware objective function is DMT loss  \cite{hu2021topology}, which helps to detect the saddle points that aid in reconstructing the topologically incorrect regions. Hu \cite{hu2022structure} computes warping errors at the homotopic level to promote topology. Recently, cbLoss \cite{shi2024centerline} was introduced to mitigate data imbalance in medical image segmentation.

\paragraph{Adaptive and Structure-aware Layers}
Dai \etal \cite{dai2017deformable} first proposed the deformable convolution networks (DCN), with its kernel learning to deform towards structures and shapes. Follow-up versions of DCN \cite{zhu2019deformable,wang2023internimage,xiong2024efficient,yeganeh2020inverse} expand this idea by adding more deformations, incorporating it into foundation models, and further improving the efficiency. With principles from DCN \cite{dai2017deformable}, \cite{dong2022deu, yang2022dcu, jin2019dunet} adapt the shape and geometry of anatomical structures dynamically. Y-Net \cite{farshad2022net} employed fast-fourier convolutions to extract spectral features from medical images. Qi \cite{qi2023dynamic} proposed snake-like kernels for deformable convolutions in dynamic snake convolutions (DSC) for topologically faithful tubular structure segmentation. However, the pre-set kernel shapes in DSC might neglect the performance while preserving the topology in other general shapes of structures. We, however, adopt a different strategy in topology preservation with an adaptive kernel; instead of pre-setting kernel shape, we aim to guide the kernel with offsets towards regions of higher topological interest.
\section{Background}
\begin{figure*}[tb]
    \centering
    \includegraphics[width=1\textwidth]{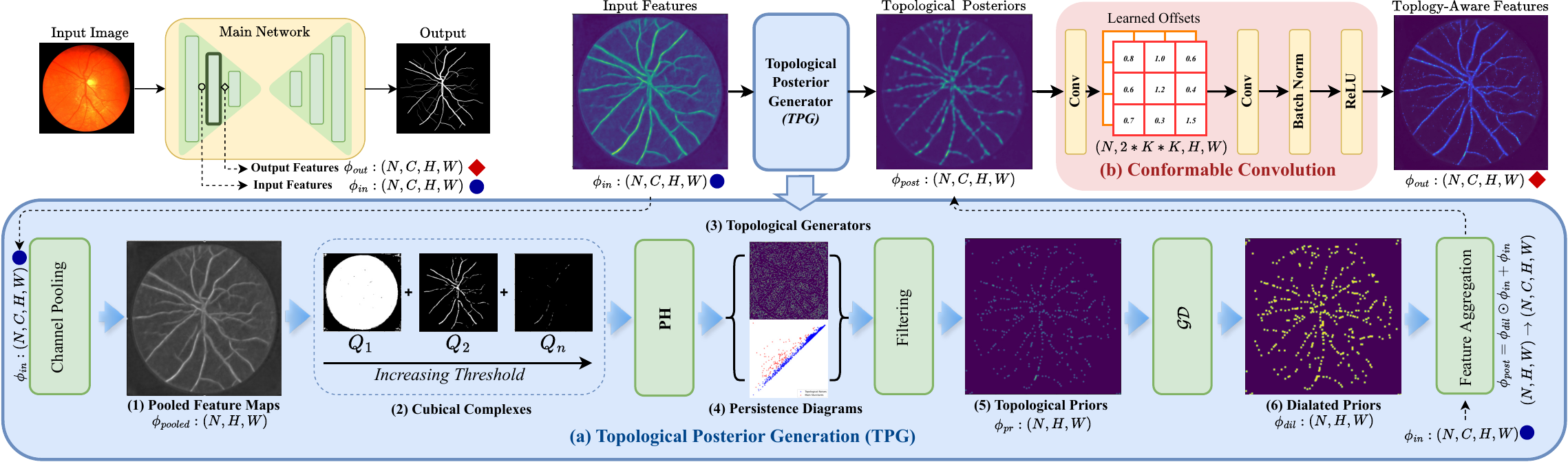}
    \caption{Our proposed layer comprises two modules: (a) \textit{Topological Posterior Generation}: receives the input feature map $\phi_{in}$ from the previous layer and generates $\phi_{post}$. (b) \textit{Conformable Convolution}: receives $\phi_{post}$, generates offsets with the first convolution layer for the adaptive kernel of the second convolution. The topology-aware features are extracted and passed through \textit{Batch Norm} and \textit{ReLU} layers. The proposed module depicts a layer that can be used at different positions in architectures such as UNet.}
    \label{fig:TPG_block}
\end{figure*}

Topological Data Analysis (TDA) \cite{wasserman2018topological} is a branch of applied mathematics focused on extracting meaningful geometric and topological features from high-dimensional, often noisy, and sparse data. Given a dataset \(X \subset \mathbb{R}^n\), TDA focuses on analyzing the topological space \((X, \Theta)\), where \(\Theta\) is an appropriate topology that captures the inherent structure of the data. Central to TDA is persistent homology, a technique that identifies and tracks topological features such as connected components, loops, and voids across multiple scales. These features are represented using simplicial complexes (\(K\)) or cubical complexes (\(Q\)), constructed from basic geometric shapes like points, lines, and triangles. These complexes serve as a bridge between the raw data (\(X\)) and its topological structure, which is quantified by homology groups such as Simplicial Complex: \( K = \bigcup_{i=0}^d \sigma_i \), where \(\sigma_i\) are simplices or Cubical Complex: \( Q = \bigcup_{i=0}^d c_i \), where \(c_i\) are cubes \cite{chazal2021introduction}. 
TDA's capacity to derive robust, qualitative insights from complex data has led to its application in various fields, including biology, neuroscience, materials science, and social network analysis \cite{moor2020topological,rieck2020uncovering}.

In 2D medical imaging, \textbf{Cubical Complexes} are particularly suitable due to the grid-like structure of the images \cite{santhirasekaram2023topology}. Formally, a cubical complex \(Q\) in a 2D binary image consists of 0-Dimensional Cubes (0-Cells): Foreground pixels, denoted as \(c_0 \in Q\) and 1-Dimensional Cubes (1-Cells): Connections between foreground pixels, denoted as \(c_1 \in Q\).
For our specific task, we focus on 0-dimensional cubes as the primary representation within the cubical complex. Persistent Homology \textbf{(PH)} tracks the evolution of these topological features (0-cells in our case) across a filtration of the cubical complex. Given a feature map $\phi$ and a threshold \(\tau\), the function \(f_{\tau}(\phi) = Q\) maps \(\phi\) to a cubical complex \(Q\). Varying the threshold \(\tau\) yields a nested sequence of cubical complexes:

\begin{equation}
     \emptyset = Q_0 \in Q_1 \in Q_2 \in ... \in Q_n = Q
     \label{eq:nested_set}
\end{equation}

\paragraph{Persistence Diagram}
As \textbf{PH} is applied, one structure will be born (appear) and dead (merged into other structures). Persistence Diagram (\textbf{PD}) documents the corresponding filtering threshold $\tau$ while a structure is born and dead. If a structure is born at $\tau_i$ and dies at $\tau_j$, the tuple $(\tau_i, \tau_j)$ would be recorded in PD. Here, we denote PD as the set containing all such tuples $\{(\tau_i, \tau_j)\}$ and a function $pers(.)$ to compute the \textit{persistence} of a tuple $(\tau_i, \tau_j)$:
\begin{equation}
     pers(\tau_i, \tau_j) = |\tau_i - \tau_j|
     \label{eq:pers}
\end{equation}

\paragraph{Topological Generators}
In 2D images, topological generators are the pixel coordinates where significant topological events (birth or death of 0-cells) occur. They visually represent the starts and ends of distinct structures in an image. \cref{fig:splatting}-(b) showcases the positions of generators in orange pixels. Since PD documents the born-and-dead tuple of filtering threshold $\tau$, we can define a function \(g\) that maps the set PD, which contains tuples of thresholds \((\tau_i, \tau_j)\), to a set $G$, which contains nested tuples of pixel coordinates \(\left( (x_i, y_i), (x_j, y_j) \right)\). So the set $G$ contains all topological generators.

\begin{equation}
   \begin{gathered}
         g: PD \mapsto G, \quad \quad g((\tau_i, \tau_j)) = \left( (x_i, y_i), (x_j, y_j) \right)
    \end{gathered}
     \label{eq:g}
\end{equation}

\begin{figure}[tb]
    \centering
    \includegraphics[width=\linewidth]{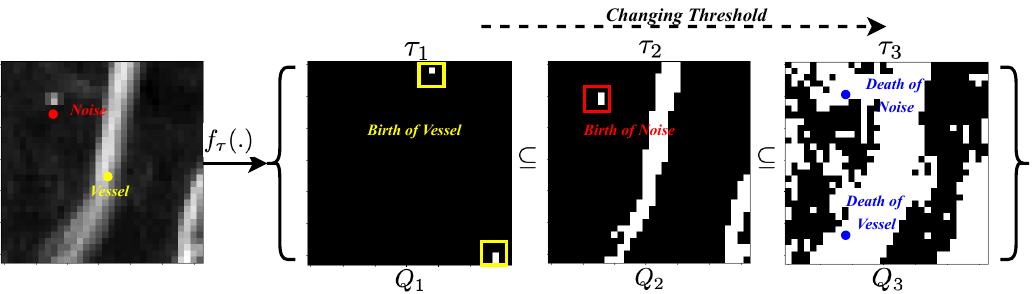} 
    \caption{An example visualization on how \textbf{PH} applies a filtering function $f_{\tau}(.)$ with changing $\tau$ ($\tau_1$, $\tau_2$, $\tau_3$) to the original image with vessel {\raisebox{0.7ex}{\colorbox{yellow}{}}} and noise {\raisebox{0.7ex}{\colorbox{red}{}}}, obtaining a nested set of cubical complexes Q ($Q_1$, $Q_2$, $Q_3$). As $\tau$ increases from $\tau_1$ to $\tau_2$, vessel {\raisebox{0.7ex}{\colorbox{yellow}{}}} is first born at $Q_1$ and noise is later born at $Q_2$ {\raisebox{0.7ex}{\colorbox{red}{}}}. Both of them {\raisebox{0.7ex}{\colorbox{blue}{}}} die at $Q_3$, as $\tau$ further raises to $\tau_3$.}
    \label{fig:background}
\end{figure}

We provide a simplified visualization of the \textbf{PH} process in \cref{fig:background}, where a nested set of Q is generated using PH. The vessel has a longer lifespan since it spans a larger range of $\tau$ compared to noise, and according to \cref{eq:pers}, the vessel has longer \textit{persistence}. This demonstrates that noise generally has shorter persistence, allowing us to filter it in our methodology.
\section{Methodology}
In this section, we present how we apply \textbf{PH} to the input feature maps and how we design our topology-guided conformable convolution layer. The methodology is divided into two subsections: Topological Posterior Generation (\textit{TPG}) (\cref{fig:TPG_block}-(a)) and Conformable Convolution (\cref{fig:TPG_block}-(b)).

Consider a semantic segmentation network $\theta$, taking an input image $I \in \mathbb{R}^{B,C',H',W'}$, and producing a predicted segmentation map $y' = \theta(I)$. Given the ground truth segmentation map $y$, the network's objective is to minimize the Dice loss \cite{milletari2016v} between $y$ and $y'$.

Our topological module can process both raw images and intermediate feature maps; therefore, it can be inserted at any intermediate layer $\theta_i$ within the network $\theta$.  When inserted as the first layer ($\theta_0$), the module operates directly on the input image $I$. For subsequent layers ($\theta_i$ ,i>0), the module processes the feature map output of the preceding layer. For notational simplicity, we refer to the input to the module generically as a feature map.

\begin{figure}[tb]
    \centering
    \includegraphics[width=\linewidth]{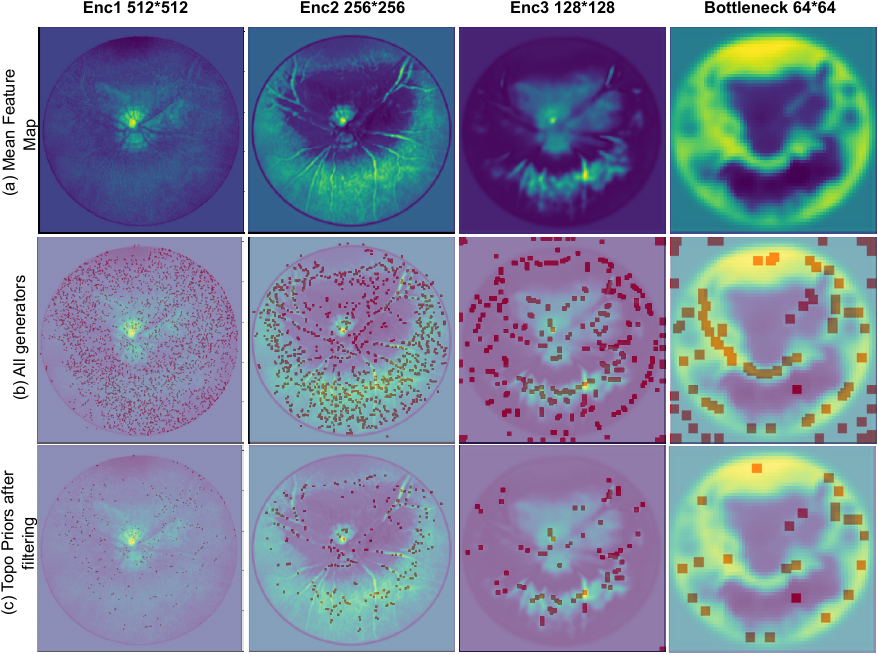}
    \caption{Visualization of Topological Priors in each layer of \textit{UNet + Conform}.}
    \label{fig:prior_vis}
\end{figure}

\subsection{Topological Posterior Generation} \label{sec:TPG}
We are given an input feature map $\phi_{in} \in \mathbb{R}^{N \times C \times H \times W}$, where $N$, $C$, $H$, and $W$ are the batch size, channels, height, and width, respectively. Our \textit{TPG} block computes a weighted prior $\phi_{pr}$ that emphasizes regions with high topological interests, then aggregates the original semantics from  $\phi_{in}$ back to the topological posterior $\phi_{post}$ which will be passed to the \textit{Conformable} block (\cref{fig:TPG_block}-(b)).

First, a channel pool layer denoted by $\psi$ is applied to $\phi_{in}$ to extract the global patterns and to reduce the channel dimensionality (\textit{cf.} \cref{fig:TPG_block}(a-1)), getting $\phi_{pooled} \in \mathbb{R}^{N \times H \times W}$:
\begin{equation}
    \phi_{pooled} = \psi(\phi_{in})  
\end{equation}

As described in the background section, \textbf{PH} is later applied to $\phi_{pooled}$ to generate a set of tuples ${\{ (\tau_i, \tau_j) \mid (\tau_i, \tau_j) \in PD \}}$, representing the birth and death times of topological features. \Cref{eq:g} then maps these tuples to a corresponding set of generators, denoted as $G$. \Cref{fig:TPG_block}-(a-3) illustrates an example of $G$ for a single $\phi_{pooled}$, highlighting the presence of numerous redundant and noisy generators. As shown in our ablation study (\cref{tabs:ablation_table}), this unfiltered noise can negatively impact the topological faithfulness of the representation.

\paragraph{Filtering Generators}
As Brunner \cite{edelsbrunner2002topological} suggests, structures with low persistence values often represent noise.  To address this, we filter the set of generators $G$, retaining only those associated with significant topological features. We denote this filtered set as $G_M$. Formally, given a pair $(\tau_i, \tau_j) \in PD$ and a filtering threshold $\tau_0$, we compute:
\begin{equation}
        \mathbb{I}(\tau_i, \tau_j) = 
        \begin{cases} 
        1 & \text{if } pers(\tau_i, \tau_j) > \tau_0, \\
        0 & \text{otherwise}.
        \end{cases}
\end{equation}
This indicator function $\mathbb{I}(.)$ allows us to construct a binary mask $M$ over the entire \textit{PD}:
\begin{equation}
    \begin{gathered}
         M =  \{ \mathbb{I}(\tau_i, \tau_j) \mid (\tau_i, \tau_j) \in PD \} , \quad \quad
         G_{M} = M \odot G
    \end{gathered}
    \label{eq:filter}
\end{equation}

Through element-wise multiplication (denoted as $\odot$), we obtain the filtered generators $G_{M}$.

\paragraph{Generating Topological Priors}
Since $G_{M}$ contains a set of coordinates of generators that mark the start and ending points of any connected components, regions with concentrated generators should be of high topological interest. The next step will be converting such coordinates into a weighted prior, encoding the topological information into the learned offsets filed, which will later be acquired by our Conformable block. Such conversion from $G_{M}$ to $\phi_{pr}$ can be easily achieved by first constructing a zero $\phi_{pr} \in \mathbb{R}^{B \times H \times W}$, then filling the (i, j) entry with ones if such entry is in $G_{M}$: 
\begin{equation}
\phi_{pr}(i, j) = 
\begin{cases} 
1 & \forall (i, j) \in G_M \\
0 & \text{otherwise}.
\end{cases}
\label{eq:binary_assignment}
\end{equation}

A visualization of topological prior at different layers of the network is provided in \cref{fig:prior_vis}.
\begin{figure}[tp]
    \centering
    \includegraphics[width=\linewidth]{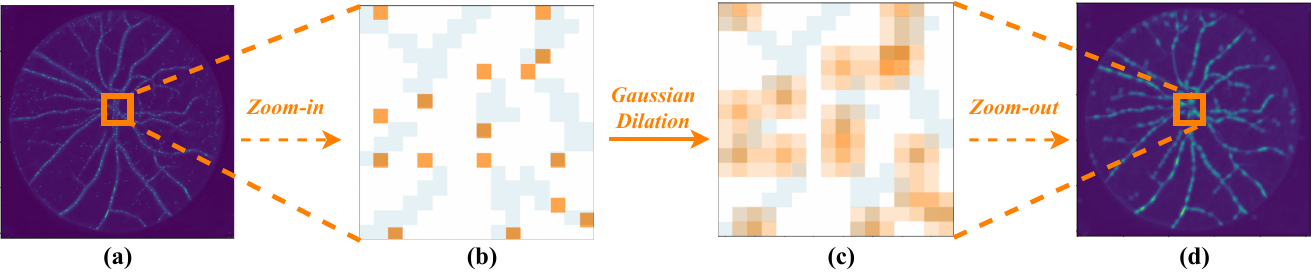} 
    \caption{Demonstration of the Gaussian dilation process on real and zoomed-in feature map: (a) $\phi_{pr}$ in a vessel feature map; (b) a zoomed-in synthetic feature map, depicting $\phi_{pr}$ emphasizing on \orange{regions of high topological interests}, (c) the effect of Gaussian dilation in dilating the \orange{topologically significant regions}; (d) the impact of Gaussian dilation on the vessel feature map.}
    \label{fig:splatting}
\end{figure}

\paragraph{Gaussian Dilation} The obtained binary $\phi_{pr}$ is indeed weighting regions with high topological interests. As depicted in \cref{fig:splatting}-(b), $\phi_{pr}$ could effectively capture the starting and ending points of a vessel and assign a weight to it. However, its pixel-wise nature makes it hard to cover all the disconnected regions. Therefore, we propose a Gaussian dilation strategy that formulates $\phi_{pr}$ as a probabilistic weighted prior. This is achieved by applying convolution to $\phi_{pr}$ with a $3\times3$ normalized Gaussian kernel denoted as $\mathcal{GD}$. We use $\ast$ to denote the convolution operator. 
\begin{equation}
    \phi_{\text{dil}} = \mathcal{GD} \ast \phi_{pr}  
\end{equation}
As shown in \cref{fig:splatting}-(c), we assign Gaussian distributions to all disconnected regions, which are of high topological interest. To visualize its effect in a real feature map, \cref{fig:splatting}-(a) and \cref{fig:splatting}-(d) show the effect after Gaussian dilation is applied. In the ablation study \cref{tabs:ablation_table}, Gaussian dilation is also justified to contribute to the topological results.

\paragraph{Topological Posterior Generation}$\phi_{pr}$ effectively augments topology significant parts. However, to prevent loss of valuable information in topological sampling, as it is shown in \cref{fig:TPG_block}-(a), the dilated prior $\phi_{dil}$ is first used to augment the topological significant parts of original input $\phi_{in}$, then it is aggregated with the  $\phi_{in}$, forming a stronger topological posterior estimation   $\phi_{post}$:

\begin{equation}
    \phi_{post} = \phi_{dil} \odot \phi_{in} + \phi_{in}
    \label{eq:aggregation}
\end{equation}

\subsection{Conformable Convolution}
Inspired by layers with an adaptive kernel design, such as the deformable convolution \cite{dai2017deformable}, we propose \textit{Conformable Convolutions}. Unlike standard convolution, convolutions with an adaptive kernel reposition convolutional kernels $w_i$ using learnable offsets $\Delta p_c$. This adaptability allows the model to better focus on contours and interconnected segments through offset convolution $g(.)$. In standard convolution, a fixed grid $R$ defines the receptive field and dilation of a kernel. The kernel elements, indexed by grid coordinates, are multiplied with corresponding pixel values from the input feature map $\phi_{in}(.)$. These products are then aggregated to produce each pixel $p$ in the output feature map $\phi_{out}(.)$, as formulated below:

\begin{equation}
  \begin{gathered}
    R = \{(-1, -1), (-1, 0), ..., (1, 1)\}, \\ \quad \quad
    \phi_{out}(p) = \sum_{p_c \in R} w_c \cdot \phi_{in}(p + p_c) 
  \end{gathered}
\end{equation}

Learnable offsets in convolution enable the kernel to sample pixel values from non-regular grid locations within the input feature map. This modulation is achieved through a set of offsets $\{\Delta p_c\}_{c=1}^C$, where $C = |R|$ represents the cardinality of the regular grid $R$ on which the kernel operates.

\begin{equation}
  \begin{gathered}
    \{\Delta p_c\}_{c=1}^C = g(\phi_{in}), \\ \quad \quad
    \phi_{out}(p) = \sum_{p_c \in R} w_c \cdot \phi_{in}(p + p_c + \Delta p_c)
  \end{gathered}
\end{equation}

The modulation of these kernels is susceptible to artifacts and high contrast inside the receptive field. In topological posterior maps, those artifacts and contrasts are suppressed by using generated birth and death points with filtration mechanisms. In this way, the adjustable convolution is still applied to the input feature maps; nonetheless, the offset adjustment is refined by topological activity regions, which introduce a new offset space with topological deformation:

\begin{equation}
  \begin{gathered}
    \{\Delta \hat{p}_c\}_{c=1}^C = g(TPG(\phi_{post})) , \\ \quad \quad
    \phi_{out}(p) = \sum_{p_c \in R} w_c \cdot \phi_{in}(p + p_c + \Delta \hat{p}_c)
  \end{gathered}
\end{equation}

\sisetup{
    table-number-alignment = center,
    round-mode = places,
    round-precision = 2,
}

\begin{table*}[tb]
\centering
\caption{\textbf{Segmentation Performance Compared to SOTA Layers with Adaptive Kernel on CHASE, HT29, and ISBI12.} The layers are inserted at the bottleneck of a UNet \cite{ronneberger2015u} model.}
\label{tabs:layers}
\renewcommand{\arraystretch}{0.9} 
\setlength{\tabcolsep}{4pt}       
\small 
\resizebox{\linewidth}{!}{
\begin{tabular}{llS[table-format=3.1]S[table-format=3.1]S[table-format=3.1]S[table-format=2.2]S[table-format=2.2]S[table-format=2.2]S[table-format=1.2]S[table-format=2.2]}
\toprule
\multicolumn{2}{c}{\textbf{Dataset}} & \multicolumn{2}{c}{\textbf{Segmentation}} & \multicolumn{6}{c}{\textbf{Continuity}} \\
\cmidrule(lr){1-2} \cmidrule(lr){3-4} \cmidrule(lr){5-10}
& & \textbf{AUC (\%)} $\uparrow$ & \textbf{Dice (\%)} $\uparrow$ & \textbf{clDice (\%)} $\uparrow$ & \boldmath{$\mathbf{error_{\beta_{0}}}$} $\downarrow$ & \boldmath{$\mathbf{error_{\beta_{1}}}$} $\downarrow$ & \boldmath{$\mathbf{error_{\chi}}$} $\downarrow$ & \textbf{ARI} $\downarrow$ & \textbf{VI} $\downarrow$ \\
\midrule

\multirow{3}{*}{HT29 \cite{carpenter2006cellprofiler,ljosa2012annotated}} 
& Deform \cite{dai2017deformable} & \textbf{99.6} & 95.8 & \textbf{93.7} & 8.20 & 13.10 & 13.30 & 0.05 & \textbf{0.19} \\ 
& DSC \cite{qi2023dynamic} & 99.4 & \textbf{95.8} & 87.6 & 8.95 & \textbf{7.83} & 20.58 & 0.06 & 0.21 \\ 
& \textbf{Conform (Ours)} & 99.1 & 94.6 & 93.1 & \textbf{5.95} & 9.6 & \textbf{6.1} & \textbf{0.04} & \textbf{0.19} \\ 
\midrule

\multirow{3}{*}{ISBI12 \cite{arganda2015crowdsourcing}} 
& Deform \cite{dai2017deformable} & 91.4 & 79.4 & 93.3 & 15.5 & 8.9 & 13.6 & 0.16 & 0.82 \\ 
& DSC \cite{qi2023dynamic} & 91.6 & 79.6 & 93.2 & 13.2 & 9.7 & 12.6 & 0.17 & 0.82 \\ 
& \textbf{Conform (Ours)} & \textbf{92.4} & \textbf{80.6} & \textbf{93.9} & \textbf{13.0} & \textbf{7.9} & \textbf{8.4} & \textbf{0.15} & \textbf{0.79} \\ 
\midrule

\multirow{3}{*}{CHASE \cite{fraz2012ensemble}} 
& Deform \cite{dai2017deformable} & 94.0 & 79.3 & 78.6 & 24.14 & 2.79 & 25.5 & 0.18 & \textbf{0.28} \\ 
& DSC \cite{qi2023dynamic} & \textbf{95.9} & 79.6 & 79.9 & 28.33 & 3.67 & 26.37 & 0.18 & 0.30 \\ 
& \textbf{Conform (Ours)} & 94.2 & \textbf{79.7} & \textbf{80.6} & \textbf{21.62} & \textbf{2.20} & \textbf{20.9} & \textbf{0.17} & \textbf{0.28} \\ 
\bottomrule
\end{tabular}
}
\end{table*}

\sisetup{
    table-number-alignment = center,
    round-mode = places,
    round-precision = 2,
}

\begin{table*}[tb]
\centering
\caption{\textbf{Segmentation Performance Compared to SOTA Segmentation Models on CHASE \cite{fraz2012ensemble}.} The best and second-best performing methods are shown in \textbf{bold} and \underline{underlined}, respectively.}
\label{tabs:chase}
\resizebox{\linewidth}{!}{
\begin{tabular}{llS[table-format=2.1]S[table-format=2.1]S[table-format=1.2]S[table-format=2.2]S[table-format=1.1]S[table-format=2.1]S[table-format=1.2]S[table-format=1.2]}
\toprule
\multicolumn{2}{c}{\textbf{Architecture}} & \multicolumn{2}{c}{\textbf{Segmentation}} & \multicolumn{5}{c}{\textbf{Continuity}} \\
\cmidrule(lr){1-2} \cmidrule(lr){3-4} \cmidrule(lr){5-9}
& & \textbf{AUC (\%)} $\uparrow$ & \textbf{Dice (\%)} $\uparrow$ & \textbf{clDice} $\uparrow$ & \boldmath{$\mathbf{error_{\beta_{0}}}$} $\downarrow$ & \boldmath{$\mathbf{error_{\beta_{1}}}$} $\downarrow$ & \boldmath{$\mathbf{error_{\chi}}$} $\downarrow$ & \textbf{ARI} $\downarrow$ & \textbf{VI} $\downarrow$ \\
\midrule

\multicolumn{9}{c}{\textbf{SOTA General Segmentation Models}} \\
\midrule
SwinUNETR \cite{hatamizadeh2021swin} & & 92.2 & 75.8 & 0.75 & 37.4 & 3.5 & 38.1 & 0.20 & 0.36 \\
SwinUNETR-V2 \cite{he2023swinunetr} & & 90.3 & 74.4 & 0.73 & 39.9 & 1.7 & 40.5 & 0.22 & 0.37 \\
FR-UNet \cite{liu2022full} & & \underline{99.1} & \underline{81.5} & 0.73 & 61.0 & 2.8 & 64.4 & \multicolumn{1}{c}{—} & \multicolumn{1}{c}{—} \\
SGL \cite{zhou2021study} & & \textbf{99.2} & \textbf{82.7} & 0.75 & 42.6 & 2.3 & 46.0 & \multicolumn{1}{c}{—} & \multicolumn{1}{c}{—} \\
\textbf{+ Conform (Ours)} & & 98.3 & 80.8 & 0.79 & 33.4 & 2.0 & 30.8 & 0.18 & \underline{0.29} \\
\midrule

\multicolumn{9}{c}{\textbf{SOTA Topological Segmentation Models}} \\
\midrule
VGN \cite{shin2019deep} & & \multicolumn{1}{c}{-} & 73.0 & 0.78 & 71.9 & 4.4 & 69.5 & \multicolumn{1}{c}{—} & \multicolumn{1}{c}{—} \\
SCOPE \cite{yeganeh2023scope} + Dice & & 95.4 & 80.0 & \underline{0.80} & 32.6 & 2.0 & 28.5 & 0.17 & \textbf{0.28} \\
\textbf{+ Conform (Ours)} & & 96.6 & 79.2 & \textbf{0.81} & 29.5 & \textbf{1.5} & 24.9 & \underline{0.15} & 0.30 \\
SCOPE \cite{yeganeh2023scope} + clDice & & 98.8 & 80.2 & \textbf{0.81} & 24.2 & \underline{1.6} & \underline{22.7} & \textbf{0.14} & 0.30 \\
\textbf{+ Conform (Ours)} & & 98.6 & 79.4 & \textbf{0.81} & \underline{21.5} & 2.1 & \textbf{19.8} & \textbf{0.14} & 0.30 \\
\midrule

\multicolumn{9}{c}{\textbf{Baseline Segmentation Models w. and w/o Conform}} \\
\midrule
UNet \cite{ronneberger2015u} & & 92.3 & 79.3 & 0.79 & 26.9 & 2.7 & 28.5 & 0.19 & 0.30 \\
\textbf{+ Conform (Ours)} & & 94.2 & 79.7 & \textbf{0.81} & 21.6 & 2.1 & \textbf{20.6} & \underline{0.17} & \textbf{0.28} \\
Y-Net \cite{farshad2022net} & & 98.0 & 78.0 & 0.76 & 27.9 & 3.1 & 24.4 & 0.18 & 0.31 \\
\textbf{+ Conform (Ours)} & & 98.7 & 80.2 & \underline{0.79} & \textbf{21.1} & 2.0 & 23.5 & \underline{0.17} & \textbf{0.28} \\
\bottomrule
\end{tabular}
}
\end{table*}

\begin{table*}[tb]
\centering
\caption{\textbf{Segmentation Performance Compared to SOTA Layers with Adaptive Kernel on CHASE, HT29, and ISBI12.} The layers are inserted at the bottleneck of a UNet \cite{ronneberger2015u} model.}
\label{tabs:layers}
\resizebox{\linewidth}{!}{
\begin{tabular}{llcccccccc}
\toprule
\multicolumn{2}{c}{\textbf{Dataset}} & \multicolumn{2}{c}{\textbf{Segmentation}} & \multicolumn{6}{c}{\textbf{Continuity}} \\
\cmidrule(lr){1-2} \cmidrule(lr){3-4} \cmidrule(lr){5-10}
& \textbf{Layer} & \textbf{AUC (\%)} $\uparrow$ & \textbf{Dice (\%)} $\uparrow$ & \textbf{clDice (\%)} $\uparrow$ & \boldmath{$\mathbf{error_{\beta_{0}}}$} $\downarrow$ & \boldmath{$\mathbf{error_{\beta_{1}}}$} $\downarrow$ & \boldmath{$\mathbf{error_{\chi}}$} $\downarrow$ & \textbf{ARI} $\downarrow$ & \textbf{VI} $\downarrow$ \\
\midrule

\multirow{3}{*}{HT29 \cite{carpenter2006cellprofiler,ljosa2012annotated}} 
& Deform \cite{dai2017deformable} & \textbf{99.6} $\pm$0.2 & 95.8 $\pm$2.1 & \textbf{93.7} $\pm$4.0 & 8.20 $\pm$3.6 & 13.10 $\pm$4.7 & 13.30 $\pm$4.2 & 0.05 $\pm$0.03 & \textbf{0.19} $\pm$0.02 \\ 
& DSC \cite{qi2023dynamic} & 99.4 $\pm$0.3 & \textbf{95.8} $\pm$2.0 & 87.6 $\pm$3.4 & 8.95 $\pm$2.8 & \textbf{7.83} $\pm$3.1 & 20.58 $\pm$7.2 & 0.06 $\pm$0.07 & 0.21 $\pm$0.01 \\ 
& \textbf{Conform (Ours)} & 99.1 $\pm$0.6 & 94.6 $\pm$1.3 & 93.1 $\pm$4.5 & \textbf{5.95} $\pm$2.4 & 9.6 $\pm$3.1 & \textbf{6.1} $\pm$2.2 & \textbf{0.04} $\pm$0.01 & \textbf{0.19} $\pm$0.06 \\ 
\midrule

\multirow{3}{*}{ISBI12 \cite{arganda2015crowdsourcing}} 
& Deform \cite{dai2017deformable} & 91.4 $\pm$0.9 & 79.4 $\pm$1.4 & 93.3 $\pm$0.8 & 15.5 $\pm$3.6 & 8.9 $\pm$3.0 & 13.6 $\pm$5.0 & 0.16 $\pm$0.1 & 0.82 $\pm$0.0 \\ 
& DSC \cite{qi2023dynamic} & 91.6 $\pm$0.2 & 79.6 $\pm$1.5 & 93.2 $\pm$0.1 & 13.2 $\pm$4.5 & 9.7 $\pm$7.0 & 12.6 $\pm$2.8 & 0.17 $\pm$0.0 & 0.82 $\pm$0.0 \\ 
& \textbf{Conform (Ours)} & \textbf{92.4} $\pm$1.5 & \textbf{80.6} $\pm$0.9 & \textbf{93.9} $\pm$0.6 & \textbf{13.0} $\pm$3.7 & \textbf{7.9} $\pm$2.9 & \textbf{8.4} $\pm$2.9 & \textbf{0.15} $\pm$0.0 & \textbf{0.79} $\pm$0.0 \\ 
\midrule

\multirow{3}{*}{CHASE \cite{fraz2012ensemble}} 
& Deform \cite{dai2017deformable} & 94.0 $\pm$0.3 & 79.3 $\pm$0.1 & 78.6 $\pm$0.3 & 24.14 $\pm$1.7 & 2.79 $\pm$0.2 & 25.5 $\pm$2.8 & 0.18 $\pm$0.00 & \textbf{0.28} $\pm$0.00 \\ 
& DSC \cite{qi2023dynamic} & \textbf{95.9} $\pm$0.2 & 79.6 $\pm$0.2 & 79.9 $\pm$0.4 & 28.33 $\pm$1.7 & 3.67 $\pm$0.5 & 26.37 $\pm$1.4 & 0.18 $\pm$0.00 & 0.30 $\pm$0.00 \\ 
& \textbf{Conform (Ours)} & 94.2 $\pm$0.2 & \textbf{79.7} $\pm$0.4 & \textbf{80.6} $\pm$0.0 & \textbf{21.62} $\pm$3.0 & \textbf{2.20} $\pm$0.4 & \textbf{20.9} $\pm$3.6 & \textbf{0.17} $\pm$0.00 & \textbf{0.28} $\pm$0.00 \\ 
\bottomrule
\end{tabular}
}
\end{table*}

\begin{figure*}[tb]
    \centering 
    \resizebox{0.98\linewidth}{!}{
    \begin{tabular}{c@{\hskip 0.15cm} c@{\hskip 0.15cm} c@{\hskip 0.15cm} c@{\hskip 0.15cm} c@{\hskip 0.15cm} c@{\hskip 0.15cm}}
         & Image & Ground Truth & Conform (Ours)  & DSC \cite{qi2023dynamic} & Deform \cite{dai2017deformable}\\
        \rotatebox{90}{\quad CHASE \cite{fraz2012ensemble}} &
        \includegraphics[width=2.8cm]{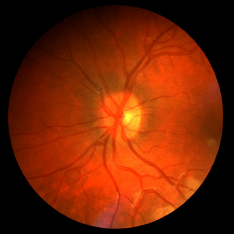}&
        \includegraphics[width=2.8cm]{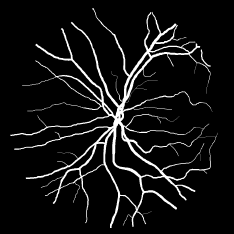}&
        \includegraphics[width=2.8cm]{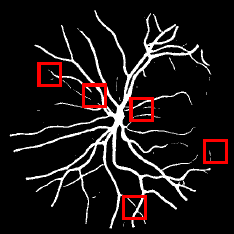}&
        \includegraphics[width=2.8cm]{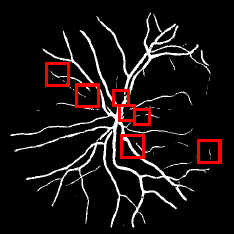}&
        \includegraphics[width=2.8cm]{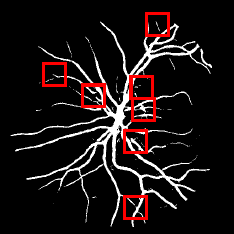}
        
        \\
        \rotatebox{90}{\quad ISBI12 \cite{arganda2015crowdsourcing}} &
        \includegraphics[width=2.8cm]{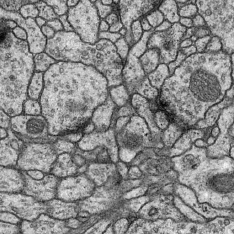}&
        \includegraphics[width=2.8cm]{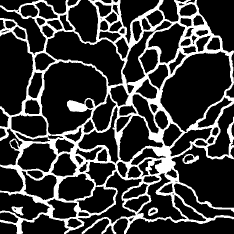}&
        \includegraphics[width=2.8cm]{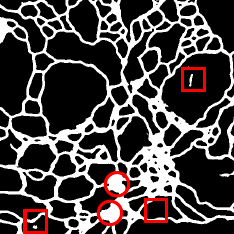}&
        \includegraphics[width=2.8cm]{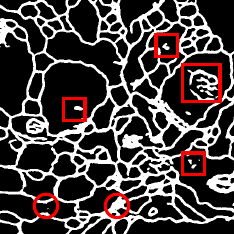}&
        \includegraphics[width=2.8cm]{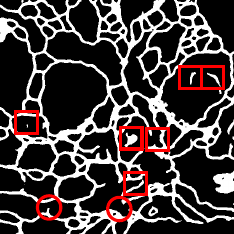}
        
        \\
        \rotatebox{90}{\quad HT29 \cite{carpenter2006cellprofiler,ljosa2012annotated}} &
        \includegraphics[width=2.8cm]{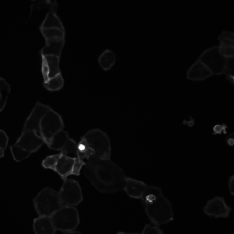}&
        \includegraphics[width=2.8cm]{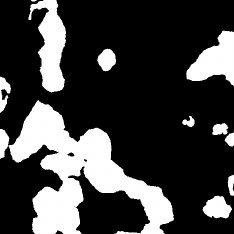}&
        \includegraphics[width=2.8cm]{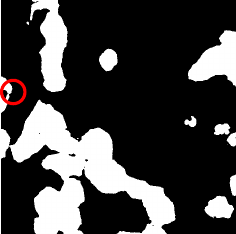}&
        \includegraphics[width=2.8cm]{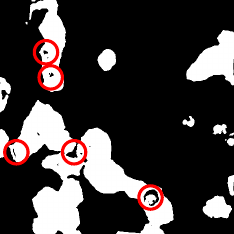}&
        \includegraphics[width=2.8cm]{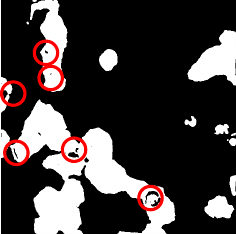}
        \\
    \end{tabular}
    }
    \caption{\textbf{Qualitative Segmentation Results corresponding to \cref{tabs:layers}}. $error_{\beta_{0}}$(highlighting disconnected components) are in \red{red squares}, while  $error_{\beta_{1}}$(highlighting holes) are in \red{red circles}.}
    \label{figs:qualitative}
\end{figure*}

\section{Experiments and Results}
In this section, we provide a comprehensive evaluation of our proposed layer for topology-aware segmentation of anatomical structures on three different medical imaging datasets: CHASE\_DB1 \cite{fraz2012ensemble}, HT29 \cite{carpenter2006cellprofiler,ljosa2012annotated} and ISBI12 \cite{arganda2015crowdsourcing}. 
First, we report the experimental setup. Then, we investigate the integration of our layer in different backbones and compare it with other state-of-the-art layers designed explicitly for modeling the geometry and topology. Then, we follow a similar strategy yet compare it against different baselines. Finally, we present an ablation study of conformable components in our layer configuration. The implementation details are reported in the supplement. 

\subsection{Experimental Setup}
\paragraph{Datasets}
We evaluate our work on three datasets with diverse topological properties, which correspond to different challenges in topology preservation. 
The \textit{ISBI12} dataset \cite{arganda2015crowdsourcing} featuring intricate network-like structures of neurons with numerous loops and connections, presents a significant challenge for preserving both \textit{0-dim} topology (\# of disconnected components) as well as \textit{1-dim} topology (\# of holes). In contrast, \textit{CHASE\_DB1} retinal vessel dataset \cite{fraz2012ensemble}, consisting of 28 images, lacks loops but exhibits complex vessel structures that demand accurate preservation of connected components (\textit{0-dim} topology). The \textit{HT29} colon cancer cell dataset from the Broad BioImage Benchmark Collection \cite{carpenter2006cellprofiler,ljosa2012annotated}, also known as \textit{BBBC}, is characterized by blob-like foreground structures with few holes, making it less sensitive to errors in \textit{1-dim} topological error, such as the $error_{\beta_{1}}$. 

\paragraph{Evaluation Metrics}
Standard classification metrics assess individual pixels within segmented regions without considering their structural relationships or connectivity. To investigate the topological properties of segmentation maps across different homology groups, as a central goal of this paper, we employ four topological and two entropy-based metrics in our evaluation. Specifically, we utilize \textit{clDice} \cite{shit2021cldice} to evaluate the center-line continuity of tubular structures. We use \textit{Betti zero} ($\beta_0$) and \textit{Betti one}($\beta_1$) \cite{vietoris1927hoheren} to count the number of connected components and independent holes, respectively. The \textit{Euler characteristic} ($\chi$) serves as a topological invariant metric, quantifying the shape of the segmentation manifold that encompasses all possible topological spaces of the segmented regions.  We employ the \textit{Adjusted Rand Index (ARI)} \cite{arganda2015crowdsourcing} to measure the similarity of randomly chosen pixel pairs belonging to the same or different segmented regions, and the \textit{Variation of Information (VI)} \cite{meilua2007comparing} to quantify the amount of information that a cluster contains about the other one. In addition to these topology-focused metrics, we report the commonly utilized pixel-wise segmentation metrics: the area under the curve (\textit{AUC}) and \textit{Dice Score} between the ground truth and predicted segmentation maps.

\subsection{Results}
\subsubsection{Comparison to Related Work}
\paragraph{Layer Comparison} We compare our proposed conformable layer to SOTA deformable layers on three medical imaging datasets by employing them in the bottleneck of the UNet \cite{ronneberger2015u} architecture. As shown in \cref{tabs:layers}, while comparing with the classic yet powerful deformable convolution layer \cite{dai2017deformable} and the SOTA Dynamic Snake Convolution (DSC) \cite{qi2023dynamic}, we observe that our conformable layer achieves best connectivity scores compared to other layers. We argue that the filtration mechanism in the topological posterior generator delineates connected and disconnected segments in feature maps, which are specifically considered to focus on those regions in convolutional deformation. This significantly enhances Betti and Euler metrics and contributes to the similarity of cluster segments (ARI and VI) and center-line connectivity (clDice) due to the amplified wholeness of anatomical structures. This shows that the conformal property of our method is capable of understanding geometry and anatomical consistency for continuity preservation but also does not sacrifice the pixel-wise results and even yields higher performance gain in dice metric.  

\paragraph{Model Comparison} We also validate the performance of our proposed layer with simple baselines compared to SOTA segmentation models in \cref{tabs:chase} on the CHASE dataset, \cref{tabs:layers} on ISBI12 and qualitatively in \cref{figs:qualitative}. In pixel-wise metrics, SGL \cite{zhou2021study} and FR-UNet \cite{liu2022full} achieve the most promising results; nevertheless, they have the difficulty to perceive inter-pixel connection and topology of segmented vessel branches. In continuity and topology preservation, SCOPE \cite{yeganeh2023scope} and conformable layer with Y-Net achieve the best results, which is also validated in our qualitative results shown in \cref{figs:qualitative}. The Conformable layer leverages topological awareness in Y-Net \cite{farshad2022net}, which provides a noticeable contribution to topological segmentation as opposed to its standard version. However, possibly due to the size of the model, there is no observable improvement in UNet \cite{ronneberger2015u}. VGN \cite{shin2019deep} is, on the contrary, liable to do over-segmentation in which curvilinear structures can be topologically segmented, yet additional isolated vessel islands would also be generated. This leads to many disconnected regions in the prediction map, thereby decreasing the dice and connectivity scores. It should be noted that although SCOPE \cite{yeganeh2023scope} achieves higher performance in some topological metrics, its architecture is designed to tackle the task at hand. Our Conform layer, on the other hand, is architecture-agnostic and can be combined with different models.

 \subsubsection{Ablation Study} 
 In this section, we ablate the effect of different components, as well as the number of Conform layers in a network. In addition, we ablate the position of inserting the Conform layer in the architecture in the supplement. 
 
 \paragraph{Effect of Different Components} To further justify our design choice of methodology in \cref{sec:TPG}, we ablate the filtration, Gaussian dilation, and feature aggregation process to learn their effects on the topological results. As shown in \cref{tabs:ablation_table}, when no filtering is applied to the generators in TPG, regions with noise would not be filtered and would be assigned a high weight in $f_{prior}$. This also leads to noise in the final prediction, causing worse topological metrics. When we remove the Gaussian dilation module in \cref{tabs:ablation_table}, the topological results also worsen. This shows that Gaussian dilation could augment the local features with topological significance, which could help the final segmentation results. At last, we also block the aggregation of input feature maps to see if the fusion of semantics from $\phi_{in}$ is really effective. After the aggregation is blocked, the \cref{eq:aggregation} is updated into:
\begin{equation}
    \phi_{post} = \phi_{dil} \odot \phi_{in}
    \label{eq:aggregation}
\end{equation}
We show that such an aggregation from $\phi_{in}$ could benefit the gradient flow and is also beneficial for the topological segmentation results in \cref{tabs:ablation_table}.

\begin{table}[tb]
\centering
\caption{\textbf{Ablation Study of Different Components on CHASE \cite{fraz2012ensemble}.} The model with all components corresponds to "UNet + Conform" in \cref{tabs:chase}. The mean and standard deviations are computed based on three different runs. $\mathcal{GD}$: Gaussian Dilation, \textbf{Fil.}: Filtration, \textbf{Aggr.}: Feature Aggregation}
\resizebox{\linewidth}{!}{
\begin{tabular}{|ccc|c c c c c c|}
\hline

Fil. & $\mathcal{GD}$ & Aggr.  & clDice $\uparrow$ & $error_{\beta_{0}}$ $\downarrow$ & $error_{\beta_{1}}$ $\downarrow$ & $error_{\chi}$ $\downarrow$ & ARI $\downarrow$ & VI $\downarrow$ \\ \hline

- & \checkmark &\checkmark & 0.79\xpm{0.00} & 32.7\xpm{1.1}  & 3.2\xpm{0.5} & 33.8\xpm{1.5}  & 0.19\xpm{0.00} & 0.30\xpm{0.02}\\   

\checkmark & - & \checkmark & 0.79\xpm{0.01} &  23.4\xpm{1.4} & 3.0\xpm{0.3}  &  23.8\xpm{2.1} & 0.19\xpm{0.01} & 0.28\xpm{0.01}\\   

\checkmark & \checkmark & - & 0.80\xpm{0.00} & 24.8\xpm{0.9} & 2.9\xpm{0.6} & 25.2\xpm{1.3} & 0.18\xpm{0.03} & 0.29\xpm{0.04}\\ 
\checkmark & \checkmark & \checkmark & \textbf{0.81}\xpm{0.00} & \textbf{21.6}\xpm{3.0}  & \textbf{2.1}\xpm{0.4} & \textbf{20.6}\xpm{3.6} & \textbf{0.17}\xpm{0.00} & \textbf{0.28}\xpm{0.00}  \\ \hline   
\end{tabular}
}
\label{tabs:ablation_table} 
\end{table}

\paragraph{Number of Conform Layers} In \cref{tabs:ablation_num_blocks}, we investigate whether increasing the number of Conform layers could lead to even better topological results. As shown in \cref{tabs:ablation_num_blocks}, we increasingly replace the standard convolution encoder blocks in the UNet \cite{ronneberger2015u} model with our Conform layer blocks. As the results indicate, a UNet with Conform layers can contribute to better topological scores. However, we notice that the topological results tend to saturate as the number of Conform blocks increases. Since one layer of Conform could already bring us satisfying results, we only choose to include one Conform block in the UNet in comparison to other architectures and methods.
\begin{table}[tb]
\centering
\caption{\textbf{Ablation Study on \# of Conform Layers on CHASE \cite{fraz2012ensemble}.} The model with "0" Conform layers denotes UNet \cite{ronneberger2015u}. Since only the best model is selected, all standard deviation errors are zero.}
\resizebox{\linewidth}{!}{
\begin{tabular}{|c|c c c c c c|}
\hline

\# of Layers  & clDice (\%) $\uparrow$ & $error_{\beta_{0}}$ $\downarrow$ & $error_{\beta_{1}}$ $\downarrow$ & $error_{\chi}$ $\downarrow$ & ARI $\downarrow$ & VI $\downarrow$ \\ \hline

0 & 79 &  26.9 & 2.7 & 28.5 & 0.19 & 0.30\\ \hline   

1  & 80 & 23.7  & 2.3 & \textbf{21.7}  & 0.17 & \textbf{0.28}\\   

2 & \textbf{81} & 23.0  &  \textbf{1.7} &  24.6 & \textbf{0.16} & \textbf{0.28} \\ 
3 & 80 & \textbf{21.8} & 2.3 & 23.6 & 0.18 & \textbf{0.28}\\ \hline
\end{tabular}
}
\label{tabs:ablation_num_blocks}
\end{table}

\section{Conclusion}
In this work, we introduced the conformable convolution layer that leverages topological priors to enhance the segmentation of intricate anatomical structures in medical images. Our novel approach incorporates a topological posterior generator (TPG) module, which identifies and prioritizes regions of high topological significance within feature maps. By integrating persistent homology, we ensure the preservation of critical topological features, such as connectivity and continuity, which are often overlooked by conventional deep learning models. Our proposed modules are designed to be architecture-agnostic, allowing seamless integration into various existing networks. Through extensive experiments on diverse medical imaging datasets, we demonstrate the effectiveness of our framework in adhering to the topology and improving segmentation performance, both quantitatively and qualitatively.

{
    \small
    \bibliographystyle{ieeenat_fullname}
    \bibliography{main}
}


\end{document}